\documentclass[twocolumn,amsmath,amssymb,10pt,aps,superscriptaddress,longbibliography]{revtex4-2} 

\usepackage{graphicx}
\usepackage{bm} % bold math
\usepackage{color}
\usepackage[breaklinks=true]{hyperref}
\usepackage{xspace}

\newcommand{\boo}{BaOsO$_3$\xspace}
\newcommand{\ttg}{t$_{2g}$\xspace}
\newcommand{\kdos}{$\rho_\mathbf{k}$\xspace}
\newcommand{\ldos}{\rho^{}_\mathrm{LDOS}\xspace}
\newcommand{\etr}{\eta_\mathrm{train}\xspace}
\newcommand{\ett}{\eta_\mathrm{test}\xspace}

\begin{document}

\title{Classification of magnetic order from electronic structure by using machine learning}

\author{Yerin \surname{Jang}}
\affiliation{Department of Physics, Chonnam National University, Gwangju 61186, Korea}
\author{Choong H. \surname{Kim}}
\email{chkim82@snu.ac.kr}
\affiliation{Center for Correlated Electron Systems, Institute for Basic Science, Seoul 08826, Korea}
\affiliation{Department of Physics and Astronomy, Seoul National University, Seoul 08826, Korea}
\author{Ara \surname{Go}}
\email[]{arago@jnu.ac.kr}
\affiliation{Department of Physics, Chonnam National University, Gwangju 61186, Korea}
  
\begin{abstract}  
Identifying the magnetic state of materials is of great interest in a wide range of applications,
but direct identification is not always straightforward due to limitations in neutron scattering experiments.
In this work, we present a machine-learning approach using decision-tree algorithms
    to identify magnetism from the spin-integrated excitation spectrum,
    such as the density of states. 
The dataset was generated by Hartree-Fock mean-field calculations of candidate antiferromagnetic orders on a Wannier Hamiltonian,
extracted from first-principle calculations targeting \boo.
Our machine learning model was trained using various types of spectral data,
including local density of states,
momentum-resolved density of states at high-symmetry points,
and the lowest excitation energies from the Fermi level.
Although the density of states shows good performance for machine learning,
    the broadening method had a significant impact on the model's performance. 
We improved the model's performance by designing the excitation energy as a feature for machine learning,
    resulting in excellent classification of antiferromagnetic order,
    even for test samples generated by different methods from the training samples used for machine learning.
\end{abstract}

\date{\today}

\maketitle

\section{Introduction}
Magnetism plays a crucial role in many physical and technological phenomena, ranging from magnetic storage devices to superconductivity.
Determining the presence of long-range magnetic ordering in materials is therefore essential for designing new functional materials with tailored magnetic properties.
%Magnetism, one of the primary characteristics of matter, is often measured as one of the first things when examining a material.
%Determining the presence of long-range magnetic order is crucial for understanding magnetism.
Neutron scattering is a powerful tool for directly determining magnetic order and is functional across a wide range of temperatures and pressures. 
However, neutron scattering experiments typically require access to specialized facilities, such as nuclear reactors or spallation sources, which can be costly. 
Additionally, it mandates a relatively large size and high-quality sample. 
The elements with high neutron absorption cross-sections also hinder clear scattering signals.

Despite the availability of direct measurement methods, the limitations mentioned make it challenging to identify magnetic order.
Therefore, it would be beneficial to have a method for determining magnetic order that is more accessible and less expensive, even if it is not as direct as neutron scattering.
For instance, specifying magnetic order based on the density of states (DOS), which can be accessed by various experimental methods, can be a compelling alternative.
In principle, magnetic orders is closely connected with the particle-hole excitation spectrum and the DOS displays distinct features of the corresponding order.
The challenge is how to extract and quantify the correlation effectively.

The recent advancement of machine learning has had a significant impact in uncovering hidden correlations in the field of condensed matter physics~%
\cite{Rosenbrock2017, Carrasquilla2017, Kelvin2017, Zhang2017, Stanev2018, Carleo2019, Ghosh2020, Lee2021b, Tsai2021}.
This technology has also been applied to the study of magnetism,
enabling for the prediction of physical quantities without the need for direct measurement or calculations,
\cite{Landrum2003, Kusne2014, Tamura2017, Miyazato2018, Nelson2019, Rhone2020, Samarakoon2020, Katsikas2021, Xie2021, Acosta2022, Chapman2022, Alidoust2022, Domina2022, Kucukbas2022} 
or probing orders from the data~\cite{Greitemann2019, Zhang2019, Shiina2020, Liu2021, Rao2021, Yu2022, Tibaldi2023}.

Motivated by the capability of machine learning to uncover complex relationships within numerical data, we explore the use of decision tree algorithms for identifying magnetic order from the density of states.
We also examine an alternative probe through momentum-resolved spectra, as the integration over the momentum space in the local density of states may mask crucial differences between various forms of magnetic order.

For the classification of magnetic order, a dataset comprising inputs and their respective magnetic order is necessary.
We selected \boo as a target system for  our machine learning study.
The polycrystalline \boo samples show metallic behavior~\cite{Shi2013, Jung2014}, but its high symmetry allows us to induce various distinct magnetic orders by lowering the symmetry.
We employ Hartree-Fock (HF) mean-field theory to generate the data sets with multiple magnetic order candidates for machine learning.
The system is ordered by local Coulomb interaction depending on antiferromagnetic order parameters we set.
The resulting DOS, momentum-resolved spectra, and antiferromagnetic orders are used to construct the data sets.

This paper is organized as follows.
In Section~\ref{sec.model}, we describe the model Hamiltonian and Hartree-Fock approximation we employed.
The data preparation for machine learning and the performance of the trained model are discussed in Section~\ref{sec.ML}.
The Section~\ref{sec.con} is devoted to conclusion and outlook.

\section{\label{sec.model}Model Hamiltonian}
\begin{figure}[tb]
	\includegraphics[width=0.98\columnwidth]{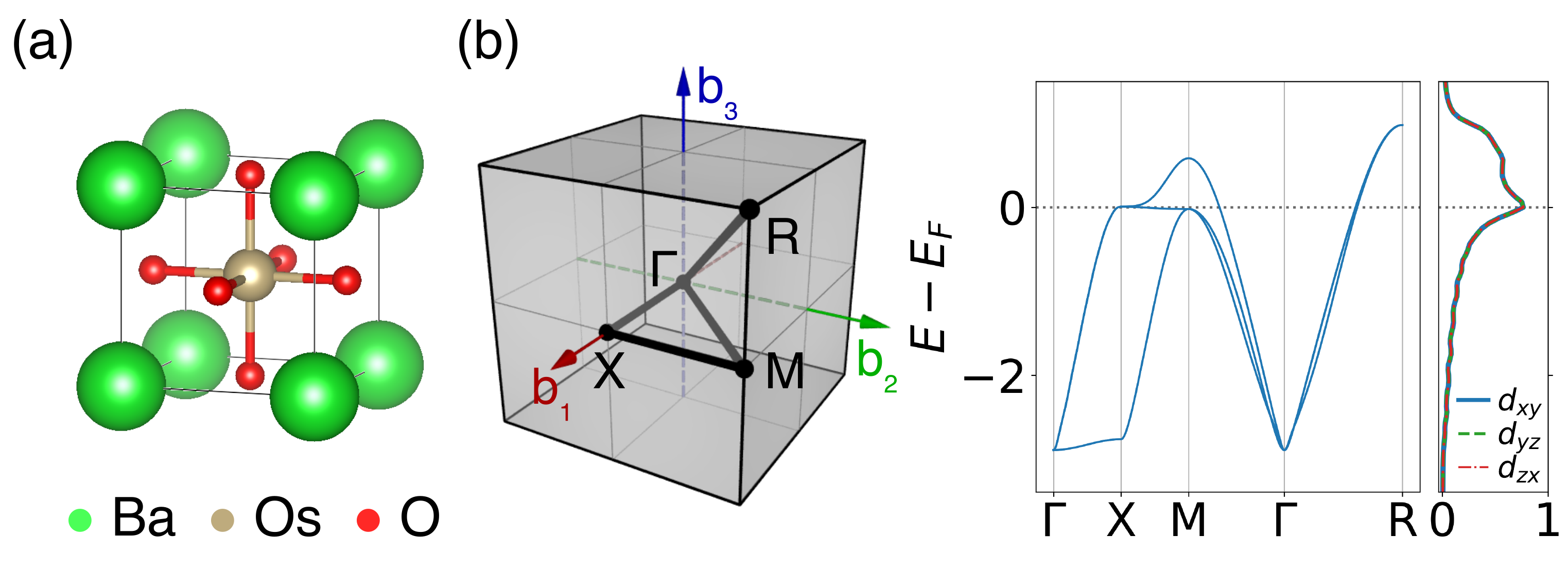}
	\caption{(a) Atomic structure of BaOsO$_3$. (b) First Brillouin zone with the path for band energies and projected density of states (PDOS) for \ttg subshells.
		In (b), the three \ttg orbitals have identical density of states due to the cubic symmetry of \boo.
	The dotted line denotes Fermi level at $N=4$.\label{fig.structure}}
\end{figure}

Figure~\ref{fig.structure} shows the unit cell and electronic structure of \boo.
The first-principles calculations were performed using density functional theory using projector augmented wave potentials within PBE exchange-correlation functional as implemented in Vienna Ab initio Simulation Package~\cite{Kresse1996,Kresse1999}.
The crystal field lifts the degeneracy of the $d$-orbitals in Os atoms,
placing the Fermi level in the $t_{2g}$ levels, which are well separated from other bands based on first-principles calculations.
We used Wannier90~\cite{Mostofi2014} to construct Maximally-Localized Wannier Functions (MLWF) based tight-binding Hamiltonians for the $t_{2g}$ bands.

\begin{figure}
	\includegraphics[width=0.82\columnwidth]{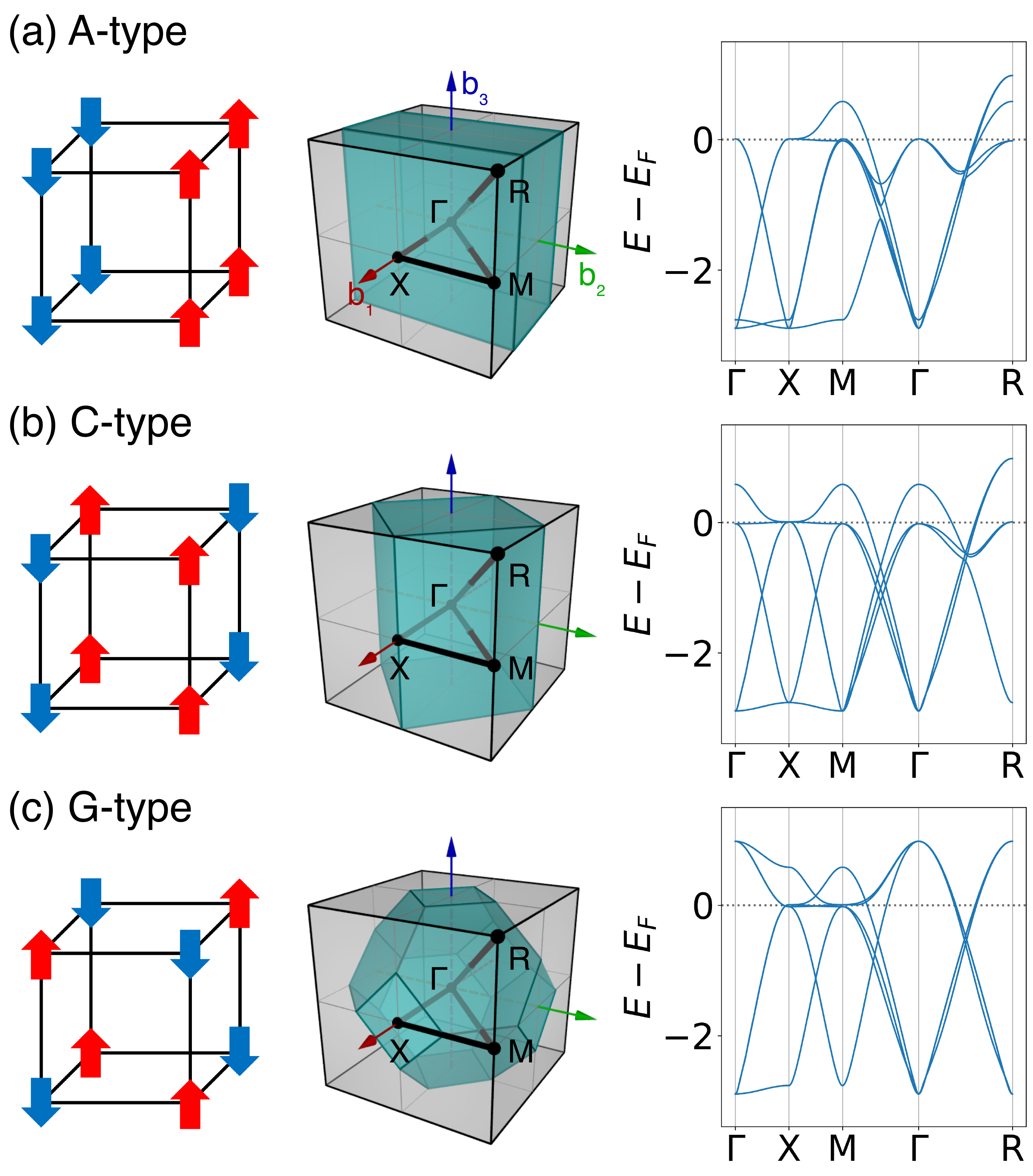}
	\caption{Schematic view and band energies with periodicity of (a) A-, (b) C- and (c) G-type antiferromagnetic order in a $2 \times 2 \times 2$ supercell. 
		The corresponding reduced Brillouin zones are represented as cyan-colored polyhedra.
		The number of bands is double that shown in Fig.~\ref{fig.structure} due to the display in the first Brillouin zone of the original lattice, ignoring the unit cell doubling for AF ordering.
		The Fermi level at $N$ = 4 is indicated by the dotted line.%
		\label{fig.fband}}
\end{figure}

The model Hamiltoinan is represented as $H = H_\mathrm{0} + H_\mathrm{int}$,
where the bilinear part is given by
\begin{align}
H_\mathrm{0} = \sum_{ij}\sum_{ll^{'}}\sum_{\sigma}t_{ij}^{ll^{'}}c^{\dagger}_{il\sigma}c_{jl^{'}\sigma},
\end{align}
where $c_{il\sigma}^{\dagger}$($c_{il\sigma}$) creates (annihilates) an electron with spin $\sigma$ in orbital $l$ at site $i$.
The hopping amplitude $t_{ij}^{ll^{'}}$ is adapted from the Wannier Hamiltonian.
The hopping parameters have ideal cubic symmetry, allowing us to induce various magnetic orders for this machine learning study.
In a $2\times 2\times 2$ supercell, three types of antiferromagnetic (AF) orders were considered: A-, C-, and G-type, as illustrated in  Fig.~\ref{fig.fband}.
In Fig.~\ref{fig.fband}, we display the three magnetic structures and their associated noninteracting band energies in the first Brillouin zone of the original lattice, which has been modified to reflect the periodicity when the respective antiferromagnetic order is stabilized.

The two-body interaction part is represented as,
\begin{align}
H_\mathrm{int} &= \frac{1}{2}U\sum_{i}\sum_{l}\sum_{\sigma} n_{il\sigma}n_{il\bar{\sigma}}\nonumber\\
&+ \frac{1}{2}U^{'}\sum_{i}\sum_{l\neq l^{'}}\sum_{\sigma} n_{i^{\vphantom{'}}l\sigma}n_{il^{'}\bar{\sigma}}\nonumber\\
&+ \frac{1}{2}U^{''}\sum_{i}\sum_{l\neq l^{'}}\sum_{\sigma} n_{i^{\vphantom{'}}l\sigma}n_{il^{'}\sigma}\nonumber\\
&+ \frac{1}{2}J\sum_{i}\sum_{l\neq l^{'}}\sum_{\sigma} c^{\dagger}_{i^{\vphantom{'}}l\sigma}c^{\dagger}_{il^{'}\bar{\sigma}}c_{i^{\vphantom{'}}l\bar{\sigma}}c_{il^{'}\sigma}\nonumber\\
&+ \frac{1}{2}J\sum_{i}\sum_{l\neq l^{'}}\sum_{\sigma}c^{\dagger}_{il\sigma}c^{\dagger}_{il\bar{\sigma}}c_{il^{'}\bar{\sigma}}c_{il^{'}\sigma},\label{eq.Hint}
\end{align}
where $n_{il\sigma}\equiv c_{il\sigma}^{\dagger}c_{il\sigma}^{\vphantom{\dagger}}$ is a number operator,
$U$ is the on-site intra-orbital Coulomb interaction and $J$ is the Hund's coupling.
We assume the rotational invariance, setting inter-orbital interaction terms, $U^\prime=U-2J$ for two different spins and $U^{\prime\prime}=U-3J$ for the same spins.
In order to stabilize an AF order, we introduce the Hartree-Fock approximation, which allows us to handle the many-body problem in Hint by using the following mean-field ansatz.
The precise wave vectors are defined as
\begin{align}
	\langle c^{\dagger}_{jl\sigma}c_{j^{'}l^{'}\sigma^{'}} \rangle = \frac{1}{2} (n_{l} + \sigma m_{l}e^{i\mathbf{q}_\alpha \cdot\mathbf{r}_{j}})\delta_{jj^{'}}\delta_{ll^{'}}\delta_{\sigma\sigma^{'}},
	\label{eq.ansatz}
\end{align}
where $n_l$($m_l$) is an electron occupancy (staggered magnetization) of orbital $l$,
$\mathbf{r}_{j}$ is the position vector of site $j$, 
$\mathbf{q}_\alpha$ is the wave vector corresponding an AF type $\alpha$=A, C, and G.
To be precise,
$\mathbf{q}^{}_\mathrm{A}=(\pi, 0, 0)$, $\mathbf{q}^{}_\mathrm{C}=(\pi, \pi, 0)$ and $\mathbf{q}^{}_\mathrm{G}=(\pi, \pi, \pi)$.
Note that the Kronecker deltas force any nonlocal, interorbital, and interspin terms to be zero in this calculation, but introducing additional off-diagonal order parameters does not alter the applicability of this machine learning study.

\begin{figure}[tb]
	\includegraphics[width=0.99\columnwidth]{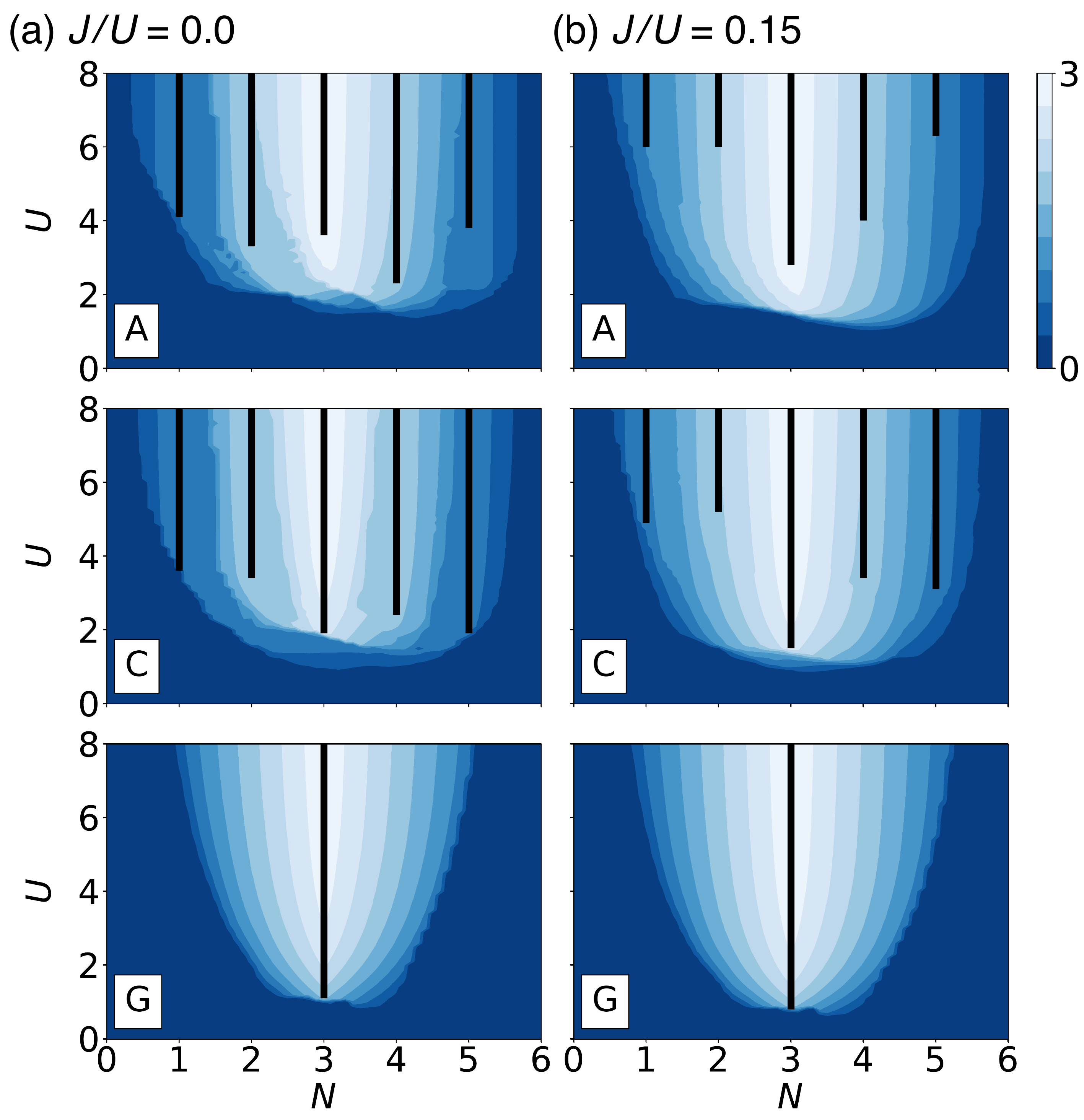}
	\caption{\label{fig.phase}Hartree-Fock phase diagrams with A-, C- and G-type antiferromagnetic ordering are shown for (a) $J/U=0.0$ and (b) $J/U=0.15$.
		The horizontal axis is the number of electrons per site and the vertical axis is the local Coulomb interaction $U$.
		The black bars indicate an insulating phase, while the color represents the staggered magnetization.
	}
\end{figure}

We perform the self-consistent calculation with the HF ansatz Eq.~(\ref{eq.ansatz}) for the three AF types
and obtain the phase diagrams as shown in Fig.~\ref{fig.phase}.
In general, the stronger interaction leads to larger staggered magnetization,
but metal-insulator transition exhibits different behavior depending on the number of electrons per site $N$ and the AF ordering.
The G-type order is metallic in a broad range of parameter space except the half-filled ($N=3$) case.
This is due to the symmetry constraint of the AF order ($\mathbf{q}_\mathrm{G}=(\pi,\pi,\pi)$),
which requires the orbital occupancies of the three \ttg orbitals to be the same.
To open a gap, the number of electrons per unit cell must be an integer, and this symmetry constraint requires it to be a multiple of three.
Therefore, the only possible case for an insulator is half-filling, as confirmed by HF calculations.

The A- and C-type antiferromagnetic orders have fewer restrictions compared to the G-type, requiring only two out of three orbital occupancies to be equal.
The different combinations of the $2+1$ occupancy splitting can result in different final solutions for a Hartree-Fock calculation.
To account for the possibility of converging to metastable states instead of the ground state,
multiple independent Hartree-Fock iterations are performed with various initial conditions
and the solution with the lowest energy is selected to construct the phase diagrams.

To increase diversity in the machine learning samples, we have included data with nonzero values of $J/U$.
The Hund's coupling has two opposing effects on the metal-insulator transition in transition metal oxides~\cite{Lee2021}.
At half-filling, it reduces the critical value of $U$,
whereas for electron fillings other than half-filling, it increases the critical $U$.
This is because the maximum total spin induced by the Hund's coupling is much larger at half-filling, making gap formation easier.
This unique behavior makes the half-filled case special,
indicating that it may be challenging to obtain accurate predictions for this case unless a sufficient number of samples for the half-filling are included in the training set.

\section{\label{sec.ML}Machine learning}
\subsection{Data preparation}

\begin{figure*}[bt]
	\includegraphics[width=0.99\textwidth]{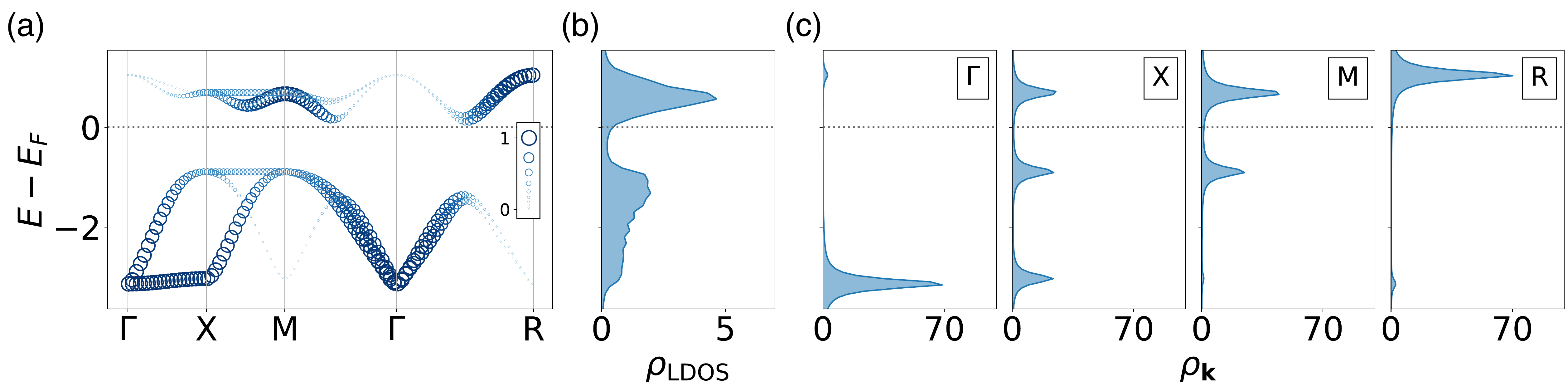}
	\caption{\label{fig.unfold}%
	(a) Unfolded band structure to restore the original periodicity for G-type order with $N=3$ and $U=2$.
	Unlike the nonmagnetic bands whose weights are identical over all momenta,
	the unfolded bands are weighted ranging from 0 to 1.
	The color and the size of circle indicate the weights.
	(b) Corresponding local density of states $\rho_\mathrm{LDOS}(\omega)$ and (c) $\mathbf{k}$-projected density of states $\rho_{\mathbf{k}}(\omega)$ at high symmetry points with a broadening factor $\eta=0.1$.
	}
\end{figure*}

We created the dataset for our machine learning study from the Hartree-Fock results discussed in the previous section.
We obtained three phase diagrams for each of the three antiferromagnetic types, with $J/U$ values of 0.0, 0.1, and 0.2.
%For each phase diagram, we collected $59\times 9$ data points, where $U$ ranges from 0 to 8 for each $N$ and $N$ varies from 0.1 to 5.9 with a step of 0.1.
For each phase diagram we collected $9\times 59$ data points, where $U$ is varied from 0 to 8 with increments of 1 and $N$ ranges from 0.1 to 5.9 with a step of 0.1.
This resulted in a total of $3 \times 3\times 9\times 59 = 4779$ samples.
The data generation took approximately 3 hours using a single CPU (Intel Xeon Platinum 8360Y with 2.40 GHz).
Still, this time could be reduced to a fraction using parallel production with multiple CPUs without sacrificing performance.

The antiferromagnetic order in HF calculations is determined by the selected HF ansatz which assumes that symmetry breaking occurs and is quantified by non-zero staggered magnetizations ($m_l$).
When $U = 0$, the $m_l$ does not contribute to the Hamiltonian and the self-consistent solution would result in $m_l$ = 0, which is equivalent to the original Hamiltonian without magnetic order.
Labeling such cases as antiferromagnetic data could negatively impact the training process.
Additionally, even if $m_l$ is not zero, extremely small values of $m_l$ only result in minimal changes to the Hamiltonian, which can confuse the machine learning model.
To improve the efficiency of the model, we included only those samples in the dataset where $m_l$ is greater than 0.1.

As this study aims to identify magnetic orders based on spin-insensitive measurements,
the input data for the machine learning should not reflect the modified periodicity in the magnetic orders,
even though the spin-integrated data contains crucial information to distinguish the AF orders.
We restore the original periodicity by applying band unfolding transformations~\cite{Boykin2005} as illustrated in Fig.~\ref{fig.unfold}(a).

In Hartree-Fock, a single particle approach, the energy axis is represented by bands as delta functions, and broadening is required to calculate the non-divergent density of states.
Despite the broadening, each peak retains the same width, as the weights of the delta functions are unity, unless two band energies are in close proximity.
This is not the case post the band unfolding process.
The local density of states (LDOS) in Fig.~\ref{fig.unfold}(b) remains unaffected by the unfolding as the weights are integrated over the Brillouin zone.
However, the momentum-resolved density of states (\kdos) in Fig.~\ref{fig.unfold}(c) shows a noticeable variation from its folded counterpart.
The energy-dependent variations in the weights of the peaks are related to the antiferromagnetic order.

The selection of optimal features is crucial because redundant features can introduce noise or bias during the learning process,
potentially leading to poor generalization or overfitting.
We test three different features in this work as described below.
We first evaluate two sets of features, one is the LDOS and the other is \kdos on the high-symmetry points ($\mathbf{k}=\Gamma$, X, M, and R).
The LDOS is calculated by integrating the \kdos over the entire Brillouin zone, as follows.
\begin{align}
	\ldos(\omega) &= \sum_{\mathbf{k}\in \mathrm{BZ}} \rho_\mathbf{k}(\omega) \nonumber \\
	&= \sum_{\mathbf{k}\in \mathrm{BZ}} \sum_n -\frac{1}{\pi} \mathrm{Im} \Big[ \frac{1}{\omega + i \eta - \varepsilon_n(\mathbf{k})} \Big],
    \label{eq.ksum}
\end{align}
where $\eta$ is a Lorentzian broadening factor and $n$ is the band index.
The calculated DOS is a continuous function of energy.
However, in order to utilize it as input features for machine learning, the DOS should be expressed as a set of numerical values.
We discretize the DOS into $N_\mathrm{bin}$ points,
\begin{align}
	\rho^{}_\mathrm{LDOS}(\omega^{}_I) = \int_{\omega^{}_I - \delta \omega}^{\omega^{}_I + \delta \omega} \rho^{}_\mathrm{LDOS}(\omega) d\omega,
	\label{eq.ldos}
\end{align}
where $I=1,2,\cdots,N_\mathrm{bin}$ are frequency indices, 
$\omega_I = -8 + I \delta \omega$,
and $\delta \omega = 16/N_\mathrm{bin}$.
The original data is generated using 1024 grid points over the energy range [-8:8],
and the features for the given $N_\mathrm{bin}$ are extracted by integrating the cubic-interpolated LDOS.
The \kdos at each high-symmetry point is also extracted in a similar manner,
but the number of bins is reduced by 1/4 to ensure a fair comparison between the two sets of features, LDOS and \kdos.

\begin{figure}[bt]
	\includegraphics[width=0.99\columnwidth]{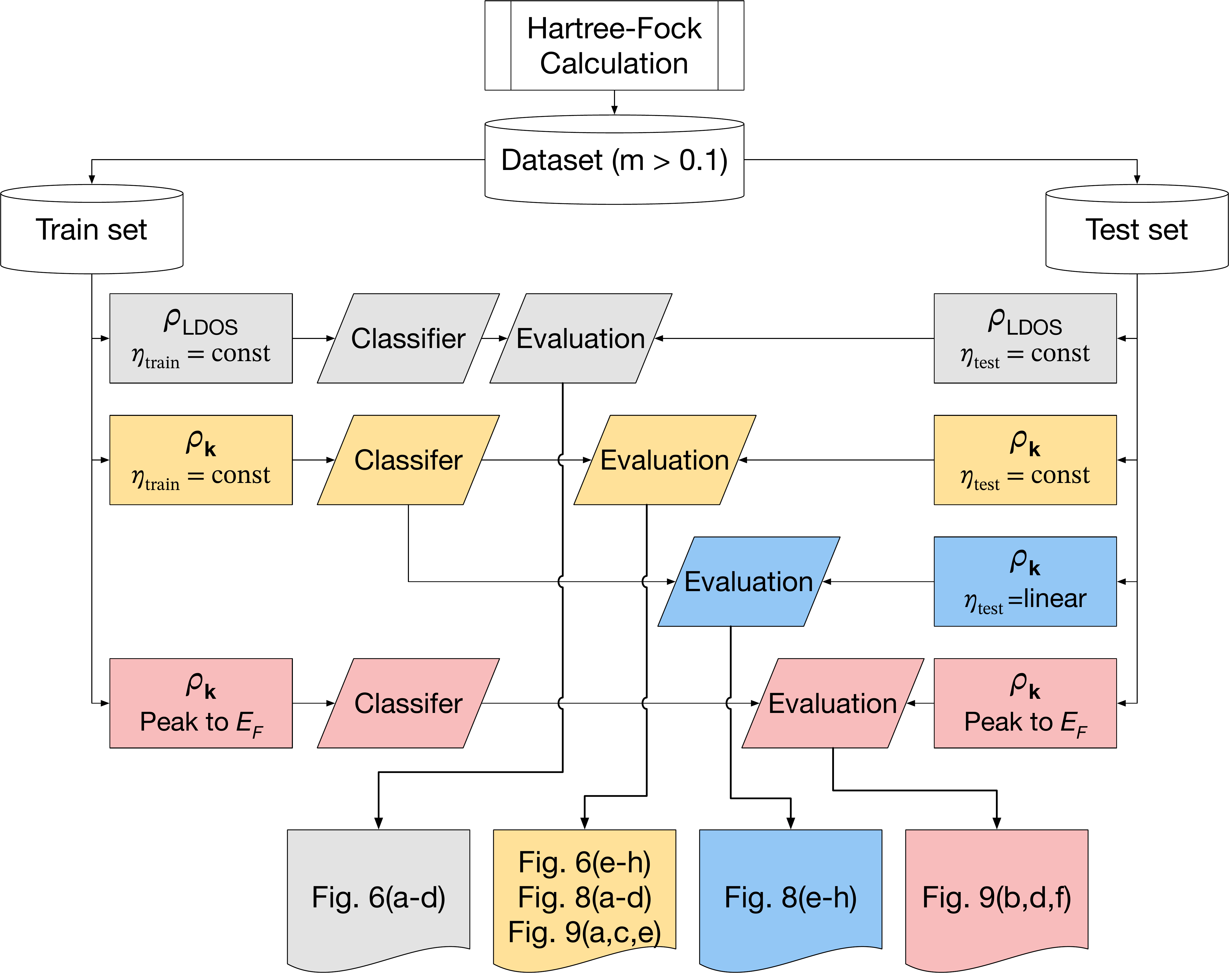}
	\caption{\label{fig.roadmap}Machine learning and data preparation roadmap.
    The combination of the features we used for training and testing is summarized with the resulting figures shown in this paper.
		}
\end{figure}

We also design the third features based on the peak structure of the \kdos feature, which will be introduced later.
An additional advantage of the third feature is its reduced dependence on the broadening scheme. Broadening can arise from various sources, including instrumental resolution, thermal fluctuations, and excitations with finite lifetimes. Given the inability to control all sources of broadening in experiments, it becomes necessary to incorporate broadening effects accurately through theoretical approaches.
The third feature is motivated by various test evaluation in Section~\ref{sec.results}.
The features we utilized for the machine learning and testing procedure, performed in Section~\ref{sec.results}, are summarized in Fig.~\ref{fig.roadmap}.

\subsection{\label{sec.results}Results of decision tree algorithms}

\begin{figure}[bt]
	\includegraphics[width=1\columnwidth]{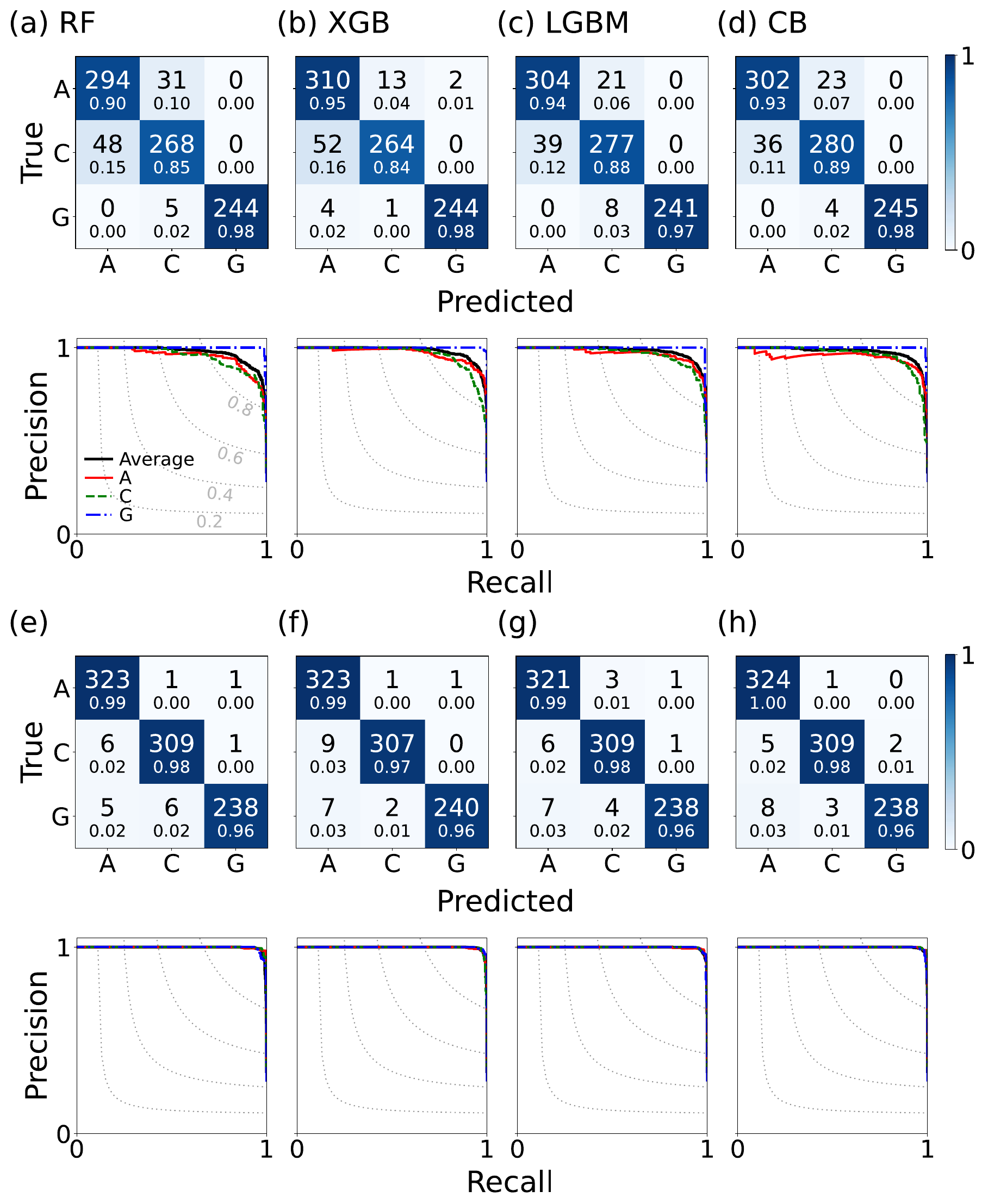}
	\caption{\label{fig.conf_dos}Confusion matrices and precision-recall curves of the machine learning models (Random Forest, XGBoost, LightGBM and CatBoost) trained on the
		(a-d) $\rho^{}_\mathrm{LDOS}$ features and
		(e-h) $\rho_{\mathbf{k}}$ at high symmetry points features with a broadening factor $\eta=0.2$.
        The color scale of the matrices represents accuracy, calculated by dividing each element of the matrix by the sum of its respective row, with the value written in small text below each element.
        In the precision-recall curves, the values for the three AF orders A, C, and G are displayed as a red solid line, a green dashed line, and a blue dash-dotted line, respectively, while their micro-average is denoted by a bold black line.
        The dotted line indicates the iso-$\mathrm{F}_1$ curve, which contains all points in the precision-recall space with the same $\mathrm{F}_1$ scores.
		}
\end{figure}

%We used decision tree ensemble algorithms available in scikit-learn~\cite{scikit-learn}, \cite{xgboost}, \cite{lightgbm} and \cite{catboost} to categorize the antiferromagnetic orders.
We used decision tree ensemble algorithms, including Random Forest, a bagging method available in scikit-learn~\cite{scikit-learn}, as well as boosting methods such as XGBoost~\cite{xgboost}, LightGBM~\cite{lightgbm} and CatBoost~\cite{catboost}.
We divided the samples into a training and a test set in a 7:3 ratio
and trained the model using Random Forest, XGBoost, LightGBM, and CatBoost algorithms.
Figure~\ref{fig.conf_dos} shows the confusion matrices and the precision-recall curves.
An element of the confusion matrix is defined as $C_{\alpha\beta}$ = (number of samples predicted as $\beta$ order while the true label is $\alpha$ in the test set).
The diagonal parts of the matrices represent cases that the trained model correctly predicts the AF labels for the test set.
The off-diagonal parts indicate the number of incorrect answers, where the column represents which labels were incorrectly assigned.

The precision-recall curve illustrates how the balance between precision and recall changes with varying thresholds. 
Precision ($P$) represents the ratio of true positives ($T_p$) to the total number of cases predicted as positive; $P = \frac{T_p}{T_p + F_p}$,
and recall ($R$) denotes the proportion of $T_p$ to the total number of actual positive samples; $R = \frac{T_p}{T_p + F_n}$.
$F_p$ and $F_n$ stand for false positives and false negatives, respectively.
$F_1$ score is the harmonic mean of precision and recall; $F_1 = 2\frac{P\times R}{P+R}$.
High precision and recall scores indicate good classification results, and the curve tends to be closer to the upper-right corner.

The performance of the models is generally good, however, errors are more frequent in cases where the filling is half or the value of $m_l$ is small. 
%The models show reasonably good performance but the majority of wrong predictions happen at half-filling or with small values of $m_l$.
The half-filled case is different from the others with non-zero Hund's coupling, making it difficult to accurately predict.
Small values of $m_l$ result in small mean-field corrections,
not providing sufficient information to distinguish different AF orders.
Because this method captures the pattern of features,
detecting weak magnetism using this method is a fundamental challenge.
Adding more half-filled samples to the training set, however, can effectively mitigate errors for gapped cases.

The accuracy of the LDOS model is relatively lower compared to the \kdos model as expected,
because the integration process during the calculation of the LDOS results in the loss of momentum-resolved information.
Despite both models having the same number of features, the performance difference suggests that feature selection is key to the success of this machine learning problem.

\begin{figure}[tb]
	\includegraphics[width=0.99\columnwidth]{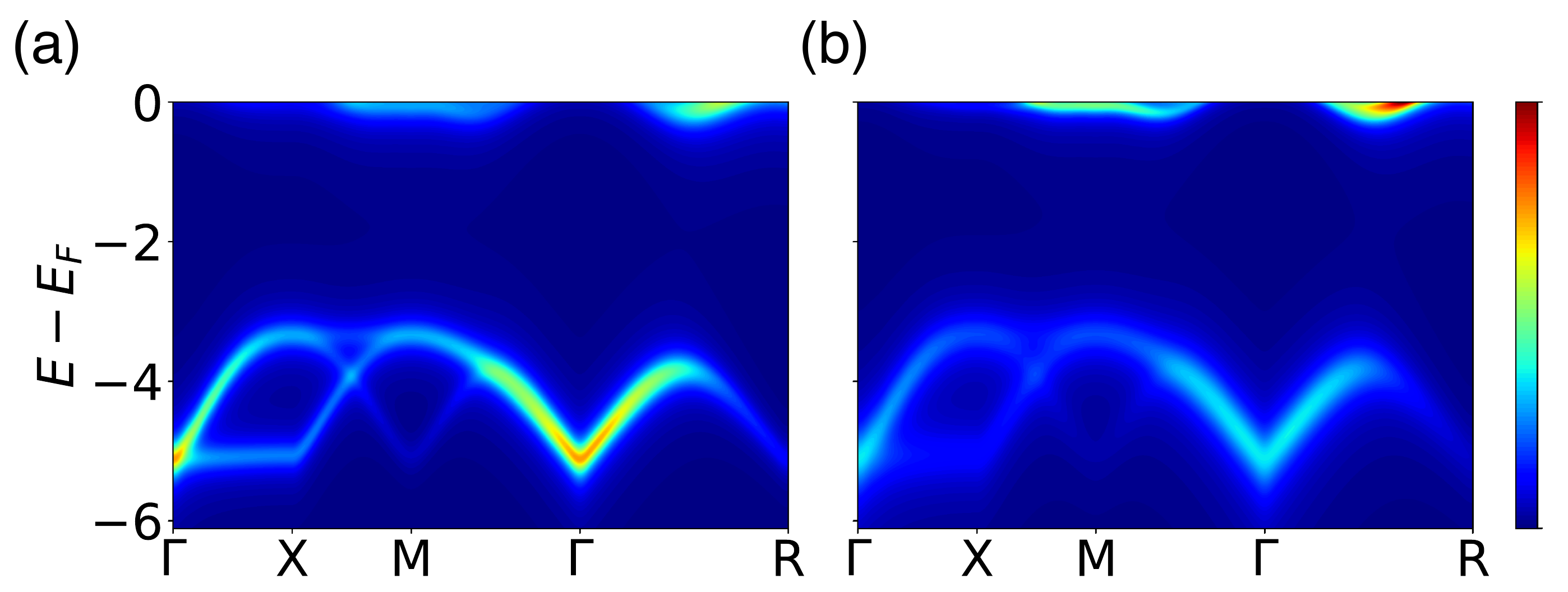}
	\caption{\label{fig.spec}Spectral weights of G-type order at $N=4$ and $U=6$ with
	(a) a constant broadening factor $\eta=0.2$ and
	(b) linearly increasing broadening as the energy lowers, $\eta(|E-E_{F}|) = 0.1 + 0.4 |E-E_{F}| / 8$. }
\end{figure}

\begin{figure}[tb]
	\includegraphics[width=1\columnwidth]{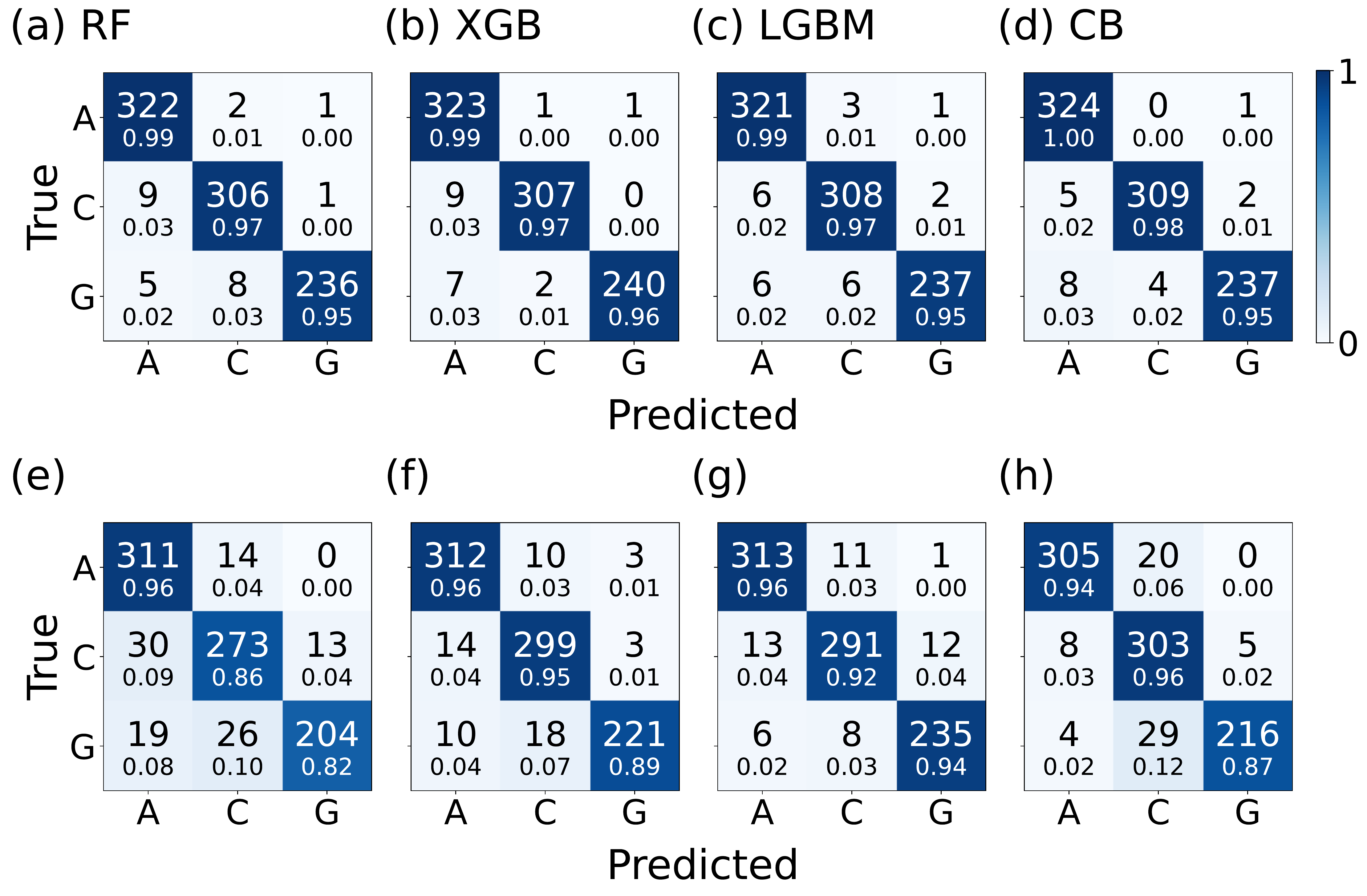}
    \caption{\label{fig.conf_arpes}Confusion matrix of machine learning classifier trained on \kdos at the high-symmetry points with $\eta_\mathrm{train}=0.2$.
    The results of Random Forest, XGBoost, LightGBM, and CatBoost classifiers are shown when the test set is broadened using
    (a-d) constant broadening ($\ett=0.2$) and (e-h) energy-dependent broadening ($\ett = 0.1 + 0.4 |E-E_{F}| / 8$).
    The color scale of the matrices represents accuracy, calculated by dividing each element of the matrix by the sum of its respective row, with the value written in small text below each element.
}
\end{figure}

The method has an advantage in that it enables feature design optimization for specific applications.
For example, angle-resolved photoemission spectroscopy (ARPES) measures hole excitation spectra, so a model that focuses on energy ranges with $E < E_F$ is necessary.
Testing the model trained within the restricted energy range would provide valuable information for analyzing the spectra.
Additionally, actual experiments can be influenced by various environmental noises, such as thermal broadening, which cannot be replicated precisely in theoretical calculations.
Thus, validating the model with different types of noise is crucial for practical applications.

Figure~\ref{fig.spec} illustrates two spectra produced from the same HF solution but with different broadening methods.
A constant broadening is applied to Fig.~\ref{fig.spec}(a),
whereas the broadening in Fig.~\ref{fig.spec}(b) increases as the energy decreases below the Fermi level as $\eta(|E-E_{F}|) = 0.1 + 0.4 |E-E_{F}| / 8$.
For validation purpose, the training set consists of DOS generated using constant broadening,
while the test set is constructed using the linearly increasing broadening as the energy lowers below the Fermi level.
The machine learning faces a more difficult situation as the test set, which is distinct from the training set, is now broadened using a different pattern from the training set.
Note that we only use the \kdos at high-symmetry points for machine learning, even though the spectra are visualized over a path connecting high-symmetry points.

We present the resulting confusion matrices in Fig.~\ref{fig.conf_arpes}.
Figure~\ref{fig.conf_arpes}(a-d) and Fig.~\ref{fig.conf_dos}(e-h) are similar in terms of accuracy,
indicating that limiting the energy range does not negatively impact the performance of the machine learning models. 
However, we observe a substantial drop in accuracy in Fig.~\ref{fig.conf_arpes}(e-h).
This decline suggests that the broadening strength significantly affects the decision trees' performance, as the trees make decisions based on numerical values that become smaller when the broadening strength $\eta$ increases.

Suppose a trained decision tree checks whether a given sample has an LDOS value above a certain threshold in a specific energy range at the root node. 
If the LDOS is larger than the threshold, it returns {\it True}; otherwise, it returns {\it False}. 
When the test set is broadened by a larger broadening factor, it reduces the height of the peak in LDOS. 
Consequently, the root node sends the sample to the {\it False} branch, and the decision tree misclassifies the sample.

\begin{figure}[tb]
	\includegraphics[width=0.99\columnwidth]{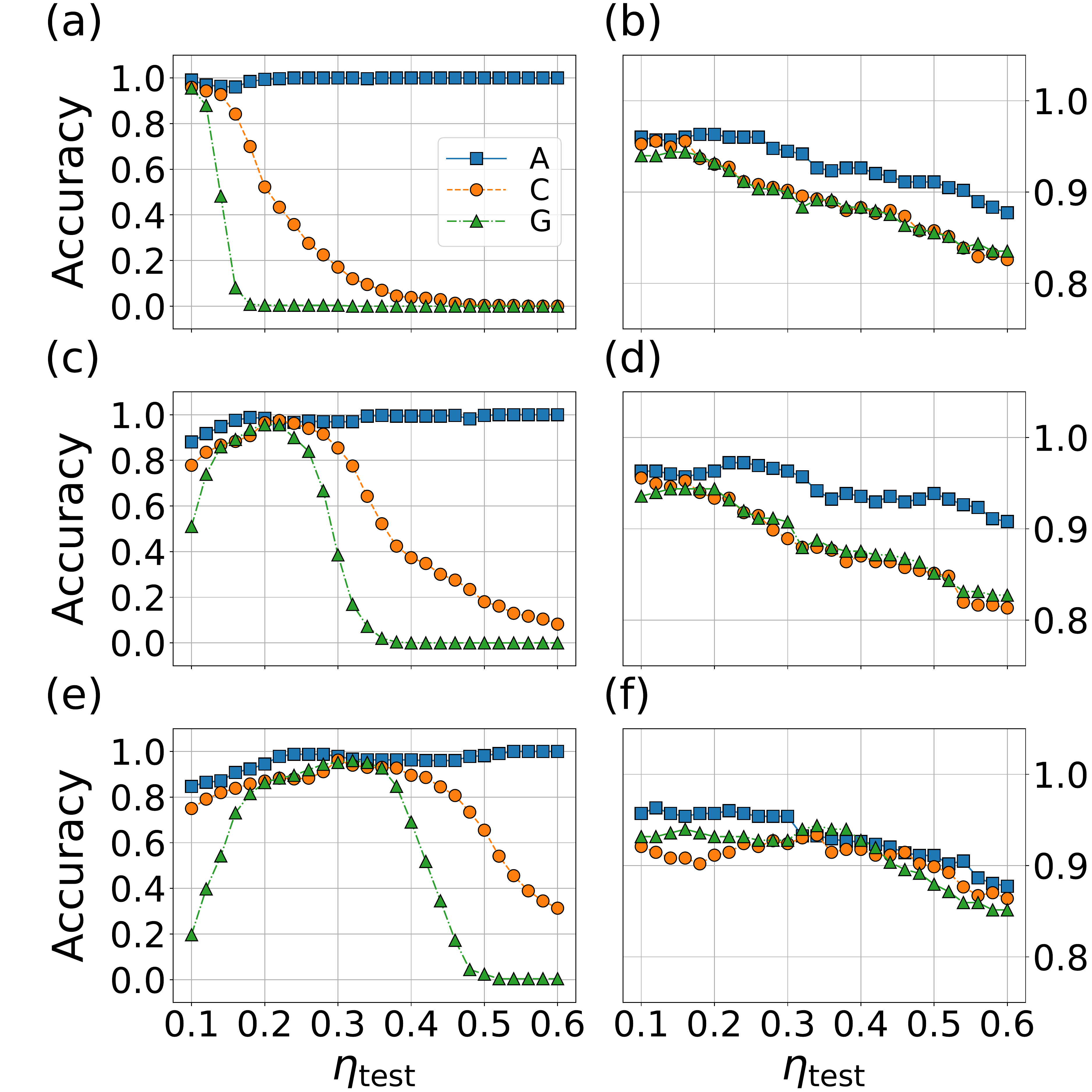}
	\caption{\label{fig.ep}Accuracies of Random Forest model trained by data with
    (a-b) $\eta_\mathrm{train}=0.1$, (c-d) $\eta_\mathrm{train}=0.2$, and (e-f) $\eta_\mathrm{train}=0.3$,
    as a function the broadening factor applied to the test set.
    The features used for the model are either \kdos (displayed in (a), (c), and (e)) or the lowest excitation energies (displayed in (b), (d), and (f)).
	}
\end{figure}

\begin{figure}[tb]
	\includegraphics[width=0.99\columnwidth]{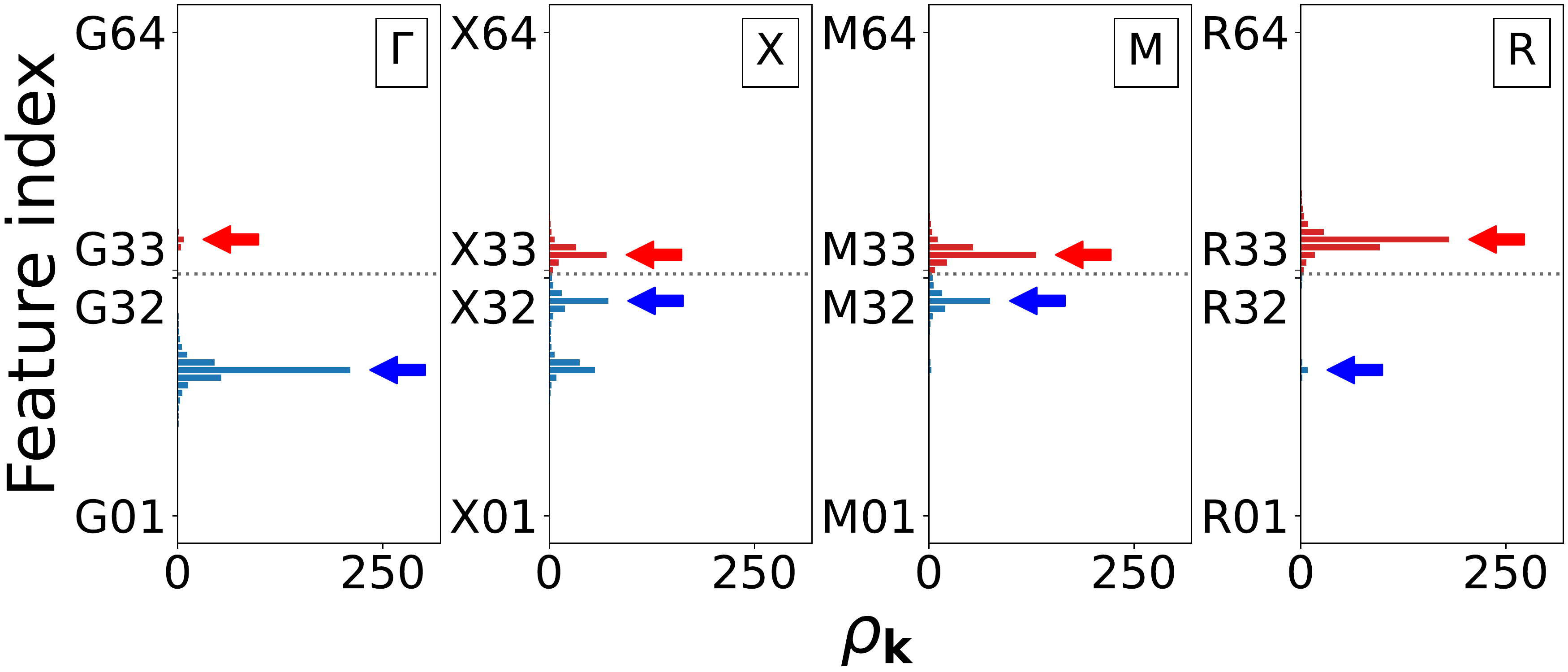}
	\caption{\label{fig.gap}An illustration of the feature section for the lowest excitations of the \kdos at high-symmetry points, for G-type order with $N=3$ and $U=2$.
    The red and blue arrows indicate the selected values as the lowest electron and hole excitation features, respectively.
	The length of bars represents the \kdos feature at each bin, extracted the HF solution shown in Fig.~\ref{fig.unfold}.
    The vertical axes are \kdos feature indices ranged from $\mathbf{k}$01 to $\mathbf{k}$64 for each $\mathbf{k}$.
    From the original \kdos features, 256 values (4$\times$64 numbers per $\mathbf{k}$),
    we choose eight features (two per $\mathbf{k}$, one positive and one negative) located above and below the Fermi level, marked by dotted gray line.
	}
\end{figure}
To investigate the impact of different broadening applied to the training and test set,
we perform a systematic cross-test of $\eta_\mathrm{train}$ and $\eta_\mathrm{test}$.
We divide the HF solutions into training and test sets,
and apply $\eta_\mathrm{train}$ and $\eta_\mathrm{test}$ to each set respectively.
The resulting accuracy is presented in Fig.~\ref{fig.ep}(a,c,e).
As expected, the best performance is achieved when $\etr = \ett$.
When $\etr$ and $\ett$ are not equal, the incorrect predictions increase, especially when $\etr$ is smaller than $\ett$.
This is because the decision trees choose a branch to follow at every nodes, based on the comparison between a certain feature and a threshold value.
A larger $\etr$ reduces these threshold values, making samples with smaller $\ett$ more likely to be classified accurately than vice versa.

Based on the results of the cross-broadening tests,
which suggest that the key aspect of the model is whether a feature (i.e. an averaged DOS within a bin) exceeds a threshold value,
we devised a compact set of features.
Despite the variation of the height and width of a peak on the energy axis as $\eta$ changes
with Lorentzian broadening in Eq.~(\ref{eq.ksum}),
the energy at which the weight reaches the local maximum remains unchanged.
This energy, originating from the corresponding band energy ($\varepsilon_n(\mathbf{k})$ in Eq.~(\ref{eq.ksum})),
represents the lowest excitation from the ground state and is measured by the distance from the Fermi level to the peak position.

Figure~\ref{fig.gap} shows the extraction of features from the original \kdos features.
For each high-symmetry point $\mathbf{k}$, the closest peak energy to the Fermi level for both electron and hole excitations.
We use the energy difference as features, where a positive value represents an electron excitation and a negative value represents a hole excitation. 
If a peak of \kdos is located on the Fermi level, the corresponding features are set to zeroes.
This feature selection significantly reduces the number of features, from 256 to 8, which is 32 times smaller than the original size,
enabling more efficient training.

The utilization of the lowest excitation energy as a feature contributes to not only efficient training but also accurate results,
as demonstrated in Fig.~\ref{fig.ep}(b,d,f).
The problematic $\eta$-dependence of the cross-$\eta$ test vanishes,
highlighting the significance of feature selection for classification based on spectral information. 
The test set encompasses both metallic and insulating solutions,
with the model tending to produce more incorrect predictions for metallic samples.
However, tests performed solely on insulating cases show over 95\% accuracy,
encouraging potential application for classifying antiferromagnetic insulators by this method.

\section{\label{sec.con}Conclusion}
In this study, we examined the application of a machine learning model for classifying antiferromagnetic (AF) orders in a model Hamiltonian targeting \boo.
The dataset was created using Hartree-Fock calculations with selected AF orders in a $2\times 2\times 2$ supercell.
Converged solutions were used to generate the local density of states (LDOS) and the momentum-resolved density of states (\kdos).
We trained the model using various features, including LDOS, \kdos, and the lowest excitations, and evaluated its ability to identify AF orders in test samples.
While both LDOS and \kdos were designed to have the same number of features, the latter demonstrated superior performance, highlighting the importance of feature selection.
The \kdos features performed well when test samples had comparable broadening levels, but different broadening methods weakened the model's performance.
In contrast, the lowest excitations, with only 8 features per sample, surpassed these limitations and exhibited excellent performance across most samples.

We considered only three types of orders in this paper, but in principle,
the approach can be extended to include more diverse types of orders.
The Hartree-Fock calculation is computationally very cheap and enables us to have a flexible design of candidate orders followed by prompt testing for desired samples.
This method will be useful for materials where conventional methods to identify magnetic orders are not applicable,
such as two-dimensional materials, non-collinear orders.

The training based on mean-field level calculations may encounter challenge in application to real materials,
in case the correlation effects lead the electronic structure to beyond the mean-field correction.
Therefore, to ensure a successful application,
it would be beneficial to validate the performance for test samples generated by methods incorporate beyond mean-field fluctuations.
For instance, identifying the AF order in dynamical mean-field calculations would be a reasonable test
as a bridge toward applications to real materials.

\begin{acknowledgments}
    This work was supported by the National Research Foundation of Korea (NRF) under Grant No. NRF2021R1C1C1010429 (Y. Jang and A. Go)
    and Institute for Basic Science under Grants No. IBS-R009-D1 (C. H. Kim).
    We thank the Center for Theoretical Physics of Complex Systems (IBS-PCS) Advanced Study Group program for their support during this collaboration.
    
\end{acknowledgments}

\bibliographystyle{naturemag}
\bibliography{magstr}

\begin{thebibliography}{10}
\expandafter\ifx\csname url\endcsname\relax
  \def\url#1{\texttt{#1}}\fi
\expandafter\ifx\csname urlprefix\endcsname\relax\def\urlprefix{URL }\fi
\providecommand{\bibinfo}[2]{#2}
\providecommand{\eprint}[2][]{\url{#2}}

\bibitem{Rosenbrock2017}
\bibinfo{author}{Rosenbrock, C.~W.}, \bibinfo{author}{Homer, E.~R.}, \bibinfo{author}{Cs{\'a}nyi, G.} \& \bibinfo{author}{Hart, G. L.~W.}
\newblock \bibinfo{title}{Discovering the building blocks of atomic systems using machine learning: application to grain boundaries}.
\newblock \emph{\bibinfo{journal}{npj Comput. Mater.}} \textbf{\bibinfo{volume}{3}}, \bibinfo{pages}{29} (\bibinfo{year}{2017}).

\bibitem{Carrasquilla2017}
\bibinfo{author}{Carrasquilla, J.} \& \bibinfo{author}{Melko, R.~G.}
\newblock \bibinfo{title}{Machine learning phases of matter}.
\newblock \emph{\bibinfo{journal}{Nat. Phys.}} \textbf{\bibinfo{volume}{13}}, \bibinfo{pages}{431--434} (\bibinfo{year}{2017}).

\bibitem{Kelvin2017}
\bibinfo{author}{Ch'ng, K.}, \bibinfo{author}{Carrasquilla, J.}, \bibinfo{author}{Melko, R.~G.} \& \bibinfo{author}{Khatami, E.}
\newblock \bibinfo{title}{Machine learning phases of strongly correlated fermions}.
\newblock \emph{\bibinfo{journal}{Phys. Rev. X}} \textbf{\bibinfo{volume}{7}}, \bibinfo{pages}{031038} (\bibinfo{year}{2017}).

\bibitem{Zhang2017}
\bibinfo{author}{Zhang, Y.} \& \bibinfo{author}{Kim, E.-A.}
\newblock \bibinfo{title}{Quantum loop topography for machine learning}.
\newblock \emph{\bibinfo{journal}{Phys. Rev. Lett.}} \textbf{\bibinfo{volume}{118}}, \bibinfo{pages}{216401} (\bibinfo{year}{2017}).

\bibitem{Stanev2018}
\bibinfo{author}{Stanev, V.} \emph{et~al.}
\newblock \bibinfo{title}{Machine learning modeling of superconducting critical temperature}.
\newblock \emph{\bibinfo{journal}{npj Comput. Mater.}} \textbf{\bibinfo{volume}{4}}, \bibinfo{pages}{29} (\bibinfo{year}{2018}).

\bibitem{Carleo2019}
\bibinfo{author}{Carleo, G.} \emph{et~al.}
\newblock \bibinfo{title}{Machine learning and the physical sciences}.
\newblock \emph{\bibinfo{journal}{Rev. Mod. Phys.}} \textbf{\bibinfo{volume}{91}}, \bibinfo{pages}{045002} (\bibinfo{year}{2019}).

\bibitem{Ghosh2020}
\bibinfo{author}{Ghosh, A.}, \bibinfo{author}{Ronning, F.}, \bibinfo{author}{Nakhmanson, S.~M.} \& \bibinfo{author}{Zhu, J.-X.}
\newblock \bibinfo{title}{Machine learning study of magnetism in uranium-based compounds}.
\newblock \emph{\bibinfo{journal}{Phys. Rev. Mater.}} \textbf{\bibinfo{volume}{4}}, \bibinfo{pages}{064414} (\bibinfo{year}{2020}).

\bibitem{Lee2021b}
\bibinfo{author}{Lee, D.}, \bibinfo{author}{You, D.}, \bibinfo{author}{Lee, D.}, \bibinfo{author}{Li, X.} \& \bibinfo{author}{Kim, S.}
\newblock \bibinfo{title}{Machine-learning-guided prediction models of critical temperature of cuprates}.
\newblock \emph{\bibinfo{journal}{J. Phys. Chem. Lett.}} \textbf{\bibinfo{volume}{12}}, \bibinfo{pages}{6211--6217} (\bibinfo{year}{2021}).

\bibitem{Tsai2021}
\bibinfo{author}{Tsai, Y.-H.} \emph{et~al.}
\newblock \bibinfo{title}{Deep learning of topological phase transitions from entanglement aspects: An unsupervised way}.
\newblock \emph{\bibinfo{journal}{Phys. Rev. B}} \textbf{\bibinfo{volume}{104}}, \bibinfo{pages}{165108} (\bibinfo{year}{2021}).

\bibitem{Landrum2003}
\bibinfo{author}{Landrum, G.~A.} \& \bibinfo{author}{Genin, H.}
\newblock \bibinfo{title}{Application of machine-learning methods to solid-state chemistry: ferromagnetism in transition metal alloys}.
\newblock \emph{\bibinfo{journal}{J. Solid State Chem.}} \textbf{\bibinfo{volume}{176}}, \bibinfo{pages}{587--593} (\bibinfo{year}{2003}).

\bibitem{Kusne2014}
\bibinfo{author}{Kusne, A.~G.} \emph{et~al.}
\newblock \bibinfo{title}{On-the-fly machine-learning for high-throughput experiments: search for rare-earth-free permanent magnets}.
\newblock \emph{\bibinfo{journal}{Sci. Rep.}} \textbf{\bibinfo{volume}{4}}, \bibinfo{pages}{6367} (\bibinfo{year}{2014}).

\bibitem{Tamura2017}
\bibinfo{author}{Tamura, R.} \& \bibinfo{author}{Hukushima, K.}
\newblock \bibinfo{title}{Method for estimating spin-spin interactions from magnetization curves}.
\newblock \emph{\bibinfo{journal}{Phys. Rev. B}} \textbf{\bibinfo{volume}{95}}, \bibinfo{pages}{064407} (\bibinfo{year}{2017}).

\bibitem{Miyazato2018}
\bibinfo{author}{Miyazato, I.}, \bibinfo{author}{Tanaka, Y.} \& \bibinfo{author}{Takahashi, K.}
\newblock \bibinfo{title}{Accelerating the discovery of hidden two-dimensional magnets using machine learning and first principle calculations}.
\newblock \emph{\bibinfo{journal}{J. Phys.: Condens. Matter}} \textbf{\bibinfo{volume}{30}}, \bibinfo{pages}{06LT01} (\bibinfo{year}{2018}).

\bibitem{Nelson2019}
\bibinfo{author}{Nelson, J.} \& \bibinfo{author}{Sanvito, S.}
\newblock \bibinfo{title}{Predicting the curie temperature of ferromagnets using machine learning}.
\newblock \emph{\bibinfo{journal}{Phys. Rev. Mater.}} \textbf{\bibinfo{volume}{3}}, \bibinfo{pages}{104405} (\bibinfo{year}{2019}).

\bibitem{Rhone2020}
\bibinfo{author}{Rhone, T.~D.} \emph{et~al.}
\newblock \bibinfo{title}{Data-driven studies of magnetic two-dimensional materials}.
\newblock \emph{\bibinfo{journal}{Sci. Rep.}} \textbf{\bibinfo{volume}{10}}, \bibinfo{pages}{15795} (\bibinfo{year}{2020}).

\bibitem{Samarakoon2020}
\bibinfo{author}{Samarakoon, A.~M.} \emph{et~al.}
\newblock \bibinfo{title}{Machine-learning-assisted insight into spin ice dy2ti2o7}.
\newblock \emph{\bibinfo{journal}{Nat. Commun.}} \textbf{\bibinfo{volume}{11}}, \bibinfo{pages}{892} (\bibinfo{year}{2020}).

\bibitem{Katsikas2021}
\bibinfo{author}{Katsikas, G.}, \bibinfo{author}{Sarafidis, C.} \& \bibinfo{author}{Kioseoglou, J.}
\newblock \bibinfo{title}{Machine learning in magnetic materials}.
\newblock \emph{\bibinfo{journal}{Phys. Status Solidi B}} \textbf{\bibinfo{volume}{258}}, \bibinfo{pages}{2000600} (\bibinfo{year}{2021}).

\bibitem{Xie2021}
\bibinfo{author}{Xie, Y.}, \bibinfo{author}{Tritsaris, G.~A.}, \bibinfo{author}{Gr{\aa}n{\"a}s, O.} \& \bibinfo{author}{Rhone, T.~D.}
\newblock \bibinfo{title}{Data-driven studies of the magnetic anisotropy of two-dimensional magnetic materials}.
\newblock \emph{\bibinfo{journal}{J. Phys. Chem. Lett.}} \textbf{\bibinfo{volume}{12}}, \bibinfo{pages}{12048--12054} (\bibinfo{year}{2021}).

\bibitem{Acosta2022}
\bibinfo{author}{Acosta, C.~M.}, \bibinfo{author}{Ogoshi, E.}, \bibinfo{author}{Souza, J.~A.} \& \bibinfo{author}{Dalpian, G.~M.}
\newblock \bibinfo{title}{Machine learning study of the magnetic ordering in 2d materials}.
\newblock \emph{\bibinfo{journal}{ACS Appl. Mater. Interfaces}} \textbf{\bibinfo{volume}{14}}, \bibinfo{pages}{9418--9432} (\bibinfo{year}{2022}).

\bibitem{Chapman2022}
\bibinfo{author}{Chapman, J. B.~J.} \& \bibinfo{author}{Ma, P.-W.}
\newblock \bibinfo{title}{A machine-learned spin-lattice potential for dynamic simulations of defective magnetic iron}.
\newblock \emph{\bibinfo{journal}{Sci. Rep.}} \textbf{\bibinfo{volume}{12}}, \bibinfo{pages}{22451} (\bibinfo{year}{2022}).

\bibitem{Alidoust2022}
\bibinfo{author}{Alidoust, M.}, \bibinfo{author}{Rothmund, E.} \& \bibinfo{author}{Akola, J.}
\newblock \bibinfo{title}{Machine-learned model hamiltonian and strength of spin--orbit interaction in strained mg2x (x = si, ge, sn, pb)}.
\newblock \emph{\bibinfo{journal}{J. Phys.: Condens. Matter}} \textbf{\bibinfo{volume}{34}}, \bibinfo{pages}{365701} (\bibinfo{year}{2022}).

\bibitem{Domina2022}
\bibinfo{author}{Domina, M.}, \bibinfo{author}{Cobelli, M.} \& \bibinfo{author}{Sanvito, S.}
\newblock \bibinfo{title}{Spectral neighbor representation for vector fields: Machine learning potentials including spin}.
\newblock \emph{\bibinfo{journal}{Phys. Rev. B}} \textbf{\bibinfo{volume}{105}}, \bibinfo{pages}{214439} (\bibinfo{year}{2022}).

\bibitem{Kucukbas2022}
\bibinfo{author}{Kucukbas, M.~E.}, \bibinfo{author}{McCann, S.} \& \bibinfo{author}{Power, S.~R.}
\newblock \bibinfo{title}{Predicting magnetic edge behavior in graphene using neural networks}.
\newblock \emph{\bibinfo{journal}{Phys. Rev. B}} \textbf{\bibinfo{volume}{106}}, \bibinfo{pages}{L081411} (\bibinfo{year}{2022}).

\bibitem{Greitemann2019}
\bibinfo{author}{Greitemann, J.}, \bibinfo{author}{Liu, K.} \& \bibinfo{author}{Pollet, L.}
\newblock \bibinfo{title}{Probing hidden spin order with interpretable machine learning}.
\newblock \emph{\bibinfo{journal}{Phys. Rev. B}} \textbf{\bibinfo{volume}{99}}, \bibinfo{pages}{060404} (\bibinfo{year}{2019}).

\bibitem{Zhang2019}
\bibinfo{author}{Zhang, Y.} \emph{et~al.}
\newblock \bibinfo{title}{Machine learning in electronic-quantum-matter imaging experiments}.
\newblock \emph{\bibinfo{journal}{Nature}} \textbf{\bibinfo{volume}{570}}, \bibinfo{pages}{484--490} (\bibinfo{year}{2019}).

\bibitem{Shiina2020}
\bibinfo{author}{Shiina, K.}, \bibinfo{author}{Mori, H.}, \bibinfo{author}{Okabe, Y.} \& \bibinfo{author}{Lee, H.~K.}
\newblock \bibinfo{title}{Machine-learning studies on spin models}.
\newblock \emph{\bibinfo{journal}{Sci. Rep.}} \textbf{\bibinfo{volume}{10}}, \bibinfo{pages}{2177} (\bibinfo{year}{2020}).

\bibitem{Liu2021}
\bibinfo{author}{Liu, K.}, \bibinfo{author}{Sadoune, N.}, \bibinfo{author}{Rao, N.}, \bibinfo{author}{Greitemann, J.} \& \bibinfo{author}{Pollet, L.}
\newblock \bibinfo{title}{Revealing the phase diagram of kitaev materials by machine learning: Cooperation and competition between spin liquids}.
\newblock \emph{\bibinfo{journal}{Phys. Rev. Res.}} \textbf{\bibinfo{volume}{3}}, \bibinfo{pages}{023016} (\bibinfo{year}{2021}).

\bibitem{Rao2021}
\bibinfo{author}{Rao, N.}, \bibinfo{author}{Liu, K.}, \bibinfo{author}{Machaczek, M.} \& \bibinfo{author}{Pollet, L.}
\newblock \bibinfo{title}{Machine-learned phase diagrams of generalized kitaev honeycomb magnets}.
\newblock \emph{\bibinfo{journal}{Phys. Rev. Res.}} \textbf{\bibinfo{volume}{3}}, \bibinfo{pages}{033223} (\bibinfo{year}{2021}).

\bibitem{Yu2022}
\bibinfo{author}{Yu, H.} \emph{et~al.}
\newblock \bibinfo{title}{Complex spin hamiltonian represented by an artificial neural network}.
\newblock \emph{\bibinfo{journal}{Phys. Rev. B}} \textbf{\bibinfo{volume}{105}}, \bibinfo{pages}{174422} (\bibinfo{year}{2022}).

\bibitem{Tibaldi2023}
\bibinfo{author}{Tibaldi, S.}, \bibinfo{author}{Magnifico, G.}, \bibinfo{author}{Vodola, D.} \& \bibinfo{author}{Ercolessi, E.}
\newblock \bibinfo{title}{{Unsupervised and supervised learning of interacting topological phases from single-particle correlation functions}}.
\newblock \emph{\bibinfo{journal}{SciPost Phys.}} \textbf{\bibinfo{volume}{14}}, \bibinfo{pages}{005} (\bibinfo{year}{2023}).

\bibitem{Shi2013}
\bibinfo{author}{Shi, Y.} \emph{et~al.}
\newblock \bibinfo{title}{High-pressure synthesis of 5d cubic perovskite baoso3 at 17 gpa: Ferromagnetic evolution over 3d to 5d series}.
\newblock \emph{\bibinfo{journal}{J. Am. Chem. Soc.}} \textbf{\bibinfo{volume}{135}}, \bibinfo{pages}{16507--16516} (\bibinfo{year}{2013}).

\bibitem{Jung2014}
\bibinfo{author}{Jung, M.-C.} \& \bibinfo{author}{Lee, K.-W.}
\newblock \bibinfo{title}{Electronic structures, magnetism, and phonon spectra in the metallic cubic perovskite ${\mathrm{baoso}}_{3}$}.
\newblock \emph{\bibinfo{journal}{Phys. Rev. B}} \textbf{\bibinfo{volume}{90}}, \bibinfo{pages}{045120} (\bibinfo{year}{2014}).

\bibitem{Kresse1996}
\bibinfo{author}{Kresse, G.} \& \bibinfo{author}{Furthm\"uller, J.}
\newblock \bibinfo{title}{Efficient iterative schemes for ab initio total-energy calculations using a plane-wave basis set}.
\newblock \emph{\bibinfo{journal}{Phys. Rev. B}} \textbf{\bibinfo{volume}{54}}, \bibinfo{pages}{11169--11186} (\bibinfo{year}{1996}).

\bibitem{Kresse1999}
\bibinfo{author}{Kresse, G.} \& \bibinfo{author}{Joubert, D.}
\newblock \bibinfo{title}{From ultrasoft pseudopotentials to the projector augmented-wave method}.
\newblock \emph{\bibinfo{journal}{Phys. Rev. B}} \textbf{\bibinfo{volume}{59}}, \bibinfo{pages}{1758--1775} (\bibinfo{year}{1999}).

\bibitem{Mostofi2014}
\bibinfo{author}{Mostofi, A.~A.} \emph{et~al.}
\newblock \bibinfo{title}{An updated version of wannier90: A tool for obtaining maximally-localised wannier functions}.
\newblock \emph{\bibinfo{journal}{Comput. Phys. Commun.}} \textbf{\bibinfo{volume}{185}}, \bibinfo{pages}{2309--2310} (\bibinfo{year}{2014}).

\bibitem{Lee2021}
\bibinfo{author}{Lee, H.~J.}, \bibinfo{author}{Kim, C.~H.} \& \bibinfo{author}{Go, A.}
\newblock \bibinfo{title}{Hund's metallicity enhanced by a van hove singularity in cubic perovskite systems}.
\newblock \emph{\bibinfo{journal}{Phys. Rev. B}} \textbf{\bibinfo{volume}{104}}, \bibinfo{pages}{165138} (\bibinfo{year}{2021}).

\bibitem{Boykin2005}
\bibinfo{author}{Boykin, T.~B.} \& \bibinfo{author}{Klimeck, G.}
\newblock \bibinfo{title}{Practical application of zone-folding concepts in tight-binding calculations}.
\newblock \emph{\bibinfo{journal}{Phys. Rev. B}} \textbf{\bibinfo{volume}{71}}, \bibinfo{pages}{115215} (\bibinfo{year}{2005}).

\bibitem{scikit-learn}
\bibinfo{author}{Pedregosa, F.} \emph{et~al.}
\newblock \bibinfo{title}{Scikit-learn: Machine learning in {P}ython}.
\newblock \emph{\bibinfo{journal}{J. Mach. Learn. Res.}} \textbf{\bibinfo{volume}{12}}, \bibinfo{pages}{2825--2830} (\bibinfo{year}{2011}).

\bibitem{xgboost}
\bibinfo{author}{Chen, T.} \& \bibinfo{author}{Guestrin, C.}
\newblock \bibinfo{title}{Xgboost: A scalable tree boosting system}.
\newblock In \emph{\bibinfo{booktitle}{Proceedings of the 22nd ACM SIGKDD International Conference on Knowledge Discovery and Data Mining}}, KDD '16, \bibinfo{pages}{785--794} (\bibinfo{publisher}{Association for Computing Machinery}, \bibinfo{address}{New York, NY, USA}, \bibinfo{year}{2016}).

\bibitem{lightgbm}
\bibinfo{author}{Ke, G.} \emph{et~al.}
\newblock \bibinfo{title}{Lightgbm: A highly efficient gradient boosting decision tree}.
\newblock In \emph{\bibinfo{booktitle}{Proceedings of the 31st International Conference on Neural Information Processing Systems}}, NIPS'17, \bibinfo{pages}{3149--3157} (\bibinfo{publisher}{Curran Associates Inc.}, \bibinfo{address}{Red Hook, NY, USA}, \bibinfo{year}{2017}).

\bibitem{catboost}
\bibinfo{author}{Prokhorenkova, L.}, \bibinfo{author}{Gusev, G.}, \bibinfo{author}{Vorobev, A.}, \bibinfo{author}{Dorogush, A.~V.} \& \bibinfo{author}{Gulin, A.}
\newblock \bibinfo{title}{Catboost: Unbiased boosting with categorical features}.
\newblock In \emph{\bibinfo{booktitle}{Proceedings of the 32nd International Conference on Neural Information Processing Systems}}, NIPS'18, \bibinfo{pages}{6639--6649} (\bibinfo{publisher}{Curran Associates Inc.}, \bibinfo{address}{Red Hook, NY, USA}, \bibinfo{year}{2018}).

\end{thebibliography}
\end{document}